\pdfoutput=1
\documentclass{JHEP3}
\usepackage{amsmath}
\usepackage{epsfig}

\def\({\left(}
\def\){\right)}
\def\<{\langle}
\def\>{\rangle}
\def\ap{\alpha}
\def\ep{\varepsilon}

\newcommand{\be}{\begin{equation}}
\newcommand{\ee}{\end{equation}}
\newcommand{\bal}{\begin{aligned}}
\newcommand{\eal}{\end{aligned}}

\newcommand{\labell}[1]{\label{#1}}

\title{All-loop Mondrian Reduction of 4-particle Amplituhedron at Positive Infinity}

\author{Junjie Rao$^{a}$\footnote{Email: jrao@aei.mpg.de}\\
{$^a$Max Planck Institute for Gravitational Physics (Albert Einstein Institute), 14476 Potsdam, Germany}}

\abstract{This article introduces a systematic framework to understand (not to derive yet)
the all-loop 4-particle amplituhedron in planar $\mathcal{N}\!=\!4$ SYM, utilizing both positivity and the
Mondrian diagrammatics. Its key idea is the simplest one so far: we can decouple one or more sets of loop variables
$(x,y,z,w)$ from the rest by just setting these variables to either zero or infinity so that their relevant positivity
conditions are trivialized, then the all-loop consistency requires that we get lower loop amplituhedra as ``residues''.
These decoupling relations connect higher loop DCI integrals with the lower ones, enabling us to identify their coefficients
starting from the 3-loop case. And surprisingly, the delicate mechanism of this process is the simple Mondrian rule
$D\!=\!X\!+\!Y$, which forces those visually non-Mondrian DCI integrals to have the correct coefficients such that the
amplituhedron can exactly reduce to the lower loop one. Examples cover all DCI integrals at $L\!=\!3,4,5,6$, especially,
the subtle 6-loop coefficients $+2$ and 0 are neatly explained in this way.}


\begin{document}
\maketitle

\section{Introduction}

The amplituhedron proposal of planar $\mathcal{N}\!=\!4$ SYM \cite{Arkani-Hamed:2013jha,Arkani-Hamed:2013kca} is a
novel reformulation which only uses positivity conditions for all physical poles to construct the amplitude or integrand.
For given $(n,k,L)$ where $n$ is the number of external particles, $(k\!+\!2)$ is the number of negative helicities and
$L$ is the loop order, the most generic loop amplituhedron is defined via
\be
Y_\ap^I=C_{\ap a}Z_a^I,~~\mathcal{L}_{(i)\ap}^I=D_{(i)\ap a}Z_a^I,
\ee
here $C_{\ap a}$ is the $(k\!\times\!n)$ positive Grassmannian encoding the tree-level information and $D_{(i)\ap a}$ is the
$(2\!\times\!n)$ positive Grassmannian with respect to the $i$-th loop, and $Z_a^I$ is the kinematical data made of $n$
generalized $(k\!+\!4)$-dimensional momentum twistors, which also obeys positivity as
\be
\<Z_{a_1}\ldots Z_{a_{k+4}}\>>0~~\textrm{for}~~a_1<\ldots<a_{k+4},
\ee
then the overall constraint is, all ordered minors of the matrix
\be
\(\begin{array}{c}
D_{(i_1)} \\
\vdots \\
D_{(i_l)} \\
C
\end{array}\)
\ee
are positive for any collection of $D$'s with $0\!\leq\!l\!\leq\!L$. Through this positive constraint
we can construct the $d\log$ form encoding logarithmic singularities of the loop amplituhedron. In practice, we also need to
know how to explicitly triangulate this geometric object, and the most recent sign-flip picture introduced in
\cite{Arkani-Hamed:2017vfh} gives a detailed prescription. However, in this work we will focus on the simplest all-loop
case of only four particles, and hence we don't need to use the sign flips.

In general, based on the definitions above, we require all physical poles to be positive:
\be
\<YZ_iZ_{i+1}Z_jZ_{j+1}\>>0,~~\<Y\mathcal{L}_{(i)}Z_jZ_{j+1}\>>0,~~\<Y\mathcal{L}_{(i)}\mathcal{L}_{(j)}\>>0,
\ee
but for the special 4-particle loop amplituhedron, there is only one sector: the MHV sector (or anti-MHV equivalently)
of $k\!=\!0$, and these constraints simplify to
\be
\<1234\>>0,~~\<\mathcal{L}_{(i)}Z_jZ_{j+1}\>>0,~~\<\mathcal{L}_{(i)}\mathcal{L}_{(j)}\>>0,
\ee
as there is no $Y$ component. We can further choose the simple positive data, and parameterize $D$'s with positive
variables $(x_i,y_i,z_i,w_i)$ as
\be
(Z_1,Z_2,Z_3,Z_4)=
\(\begin{array}{cccc}
1~ & 0~ & 0~ & 0 \\
0~ & 1~ & 0~ & 0 \\
0~ & 0~ & 1~ & 0 \\
0~ & 0~ & 0~ & 1
\end{array}\),~~
D_{(i)}=
\(\begin{array}{cccc}
1~ & x_i~ & 0~ & -w_i \\
0~ & y_i~ & 1~ & z_i
\end{array}\),
\ee
such that $\<\mathcal{L}_{(i)}Z_jZ_{j+1}\>\!>\!0$ can be trivialized as (note the twisted cyclicity $Z_{4+1}\!=\!-Z_1$)
\be
\<\mathcal{L}_{(i)}12\>=w_i,~\<\mathcal{L}_{(i)}23\>=z_i,~\<\mathcal{L}_{(i)}34\>=y_i,~\<\mathcal{L}_{(i)}14\>=x_i,
\ee
and the only nontrivial constraint is
\be
\<\mathcal{L}_{(i)}\mathcal{L}_{(j)}\>=\textrm{det}
\(\begin{array}{cccc}
1~ & x_i~ & 0~ & -w_i \\
0~ & y_i~ & 1~ & z_i \\
1~ & x_j~ & 0~ & -w_j \\
0~ & y_j~ & 1~ & z_j
\end{array}\)>0.
\ee
In summary, for the 4-particle amplituhedron or integrand at $L$-loop order, we have the mutual positivity condition for
any two sets of loop variables labeled by $i,j\!=\!1,\ldots,L$ as
\be
\<\mathcal{L}_{(i)}\mathcal{L}_{(j)}\>\equiv D_{ij}=(x_j-x_i)(z_i-z_j)+(y_j-y_i)(w_i-w_j)>0,
\ee
where positive variables $x_i,y_i,z_i,w_i$ and $D_{ij}$ are all possible physical poles.
Though the dominating principle is simple and symmetric up to all loops,
as the loop order increases, its calculational complexity grows explosively due to the highly nontrivial
intertwining of all $L(L\!-\!1)/2$ positivity conditions. Therefore it is more practical to seek
new perspectives or techniques other than confronting the direct calculation. This, however, does not imply
the direct calculation is impossible, as a better interpretation might redefine
the problem so that the meaning of ``direct'' is more trivialized. This work shows how a simpler problem
got complicated, then returns to its plain form after we switch to the correct perspective extracted from all the
previous clues. So it is natural to expect the ultimate solution of 4-particle amplituhedron turns out to be even simpler,
and hidden elegant patterns like the Mondrian story await to be discovered.

The most recent progress includes the direct calculation of the 3-loop case \cite{Rao:2017fqc}, the all-loop Mondrian
diagrammatics \cite{An:2017tbf} for a subset of dual conformally invariant (DCI) loop integrals of which
pole structures can be Mondrianized, and the positive cuts \cite{Rao:2018uta} as a simplified approach to identify
coefficients of a given basis of DCI integrals. This work continues to explore the 4-particle amplituhedron at
higher loop orders, as we will introduce a new systematic framework to more clearly integrate positivity with the
Mondrian diagrammatics. Note that, though we will use some terminologies already involved in the previous works,
such as ``Mondrian diagrammatics'', in this new setting their meanings are slightly different. To make the current work
as self-contained as possible, we will redefine the frequently used terms so it is not necessary to
recall them back in \cite{Rao:2017fqc,An:2017tbf,Rao:2018uta}.

The key idea here is straightforward: we can decouple one or more sets of loop variables $(x_i,y_i,z_i,w_i)$
from the rest by setting them to either zero or infinity to trivialize the relevant positivity conditions,
then the all-loop consistency requires that we get lower loop amplituhedra as residues.
Using these decoupling relations to connect DCI integrals of different loop orders, we can identify coefficients of
DCI integrals at $L\!\geq\!4$ starting from the 3-loop case.
And it is the simple Mondrian rule $D\!=\!X\!+\!Y$ that accounts for this delicate mechanism and forces those
visually non-Mondrian DCI integrals to have the correct coefficients.

To begin to rebuild everything, forgetting all later advances, we can return to the original definition of this problem
\cite{Arkani-Hamed:2013jha,Arkani-Hamed:2013kca}. First let's introduce a convenient convention for the following
derivations: we will use the \textit{dimensionless ratio} as the integrand, for example, in the 2-loop integral
\cite{Arkani-Hamed:2013kca}
\be
\int\frac{dx_1}{x_1}\frac{dy_1}{y_1}\frac{dz_1}{z_1}\frac{dw_1}{w_1}
\frac{dx_2}{x_2}\frac{dy_2}{y_2}\frac{dz_2}{z_2}\frac{dw_2}{w_2}\,R~~\textrm{where}~~
R=\frac{x_2z_1+x_1z_2+y_2w_1+y_1w_2}{D_{12}},
\ee
$R$ is the integrand we will extensively manipulate. In other words, the full integral is made up of the $d\log$ measure
of all positive variables and this ratio $R$. In particular, the 1-loop integrand is trivially 1 as there is no mutual
positivity to be imposed. With this convention, when the integral is evaluated at either zero or infinity with respect to
some variables, there is no extra factor to be added to the residual integrand, and the $d\log$ measure of those
unfixed variables can be dropped for convenience.

Then what use does this residue evaluation at zero or infinity have, which seems trivial compared to the positive cuts
\cite{Rao:2018uta}? If we simply set $x_j\!\to\!\infty$ and $z_j\!=\!0$ in
\be
D_{ij}=(x_j-x_i)(z_i-z_j)+(y_j-y_i)(w_i-w_j)\to\infty\,z_i+(y_j-y_i)(w_i-w_j),
\ee
$D_{ij}$ becomes trivially positive, since the positivity of $z_i$ is magnified by a positive infinity factor.

\begin{figure}
\begin{center}
\includegraphics[width=0.7\textwidth]{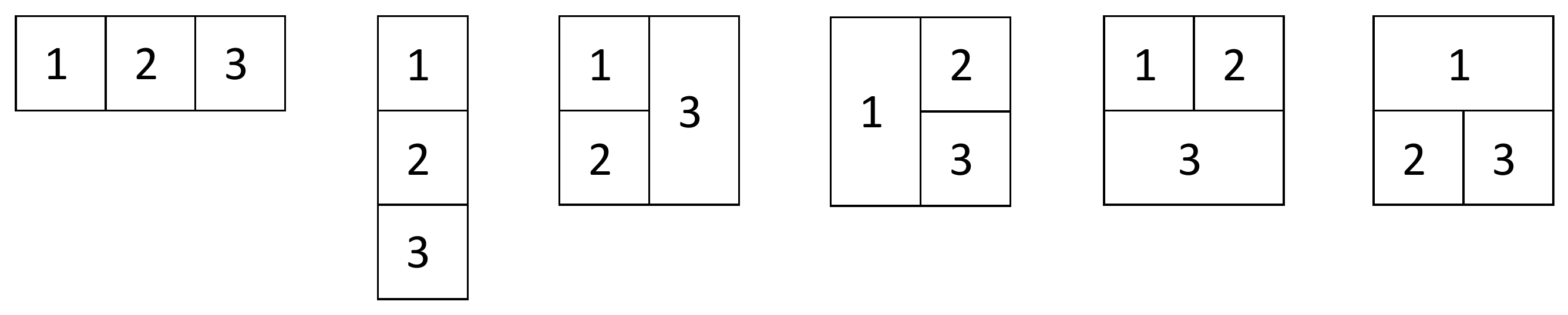}
\caption{Diagrams of DCI integrals corresponding to the first six terms in $R_3$. In our convention, faces $x,z,y,w$
locate at the left, right, top and bottom of the diagram respectively.} \label{fig-0}
\end{center}
\end{figure}

\begin{figure}
\begin{center}
\includegraphics[width=0.85\textwidth]{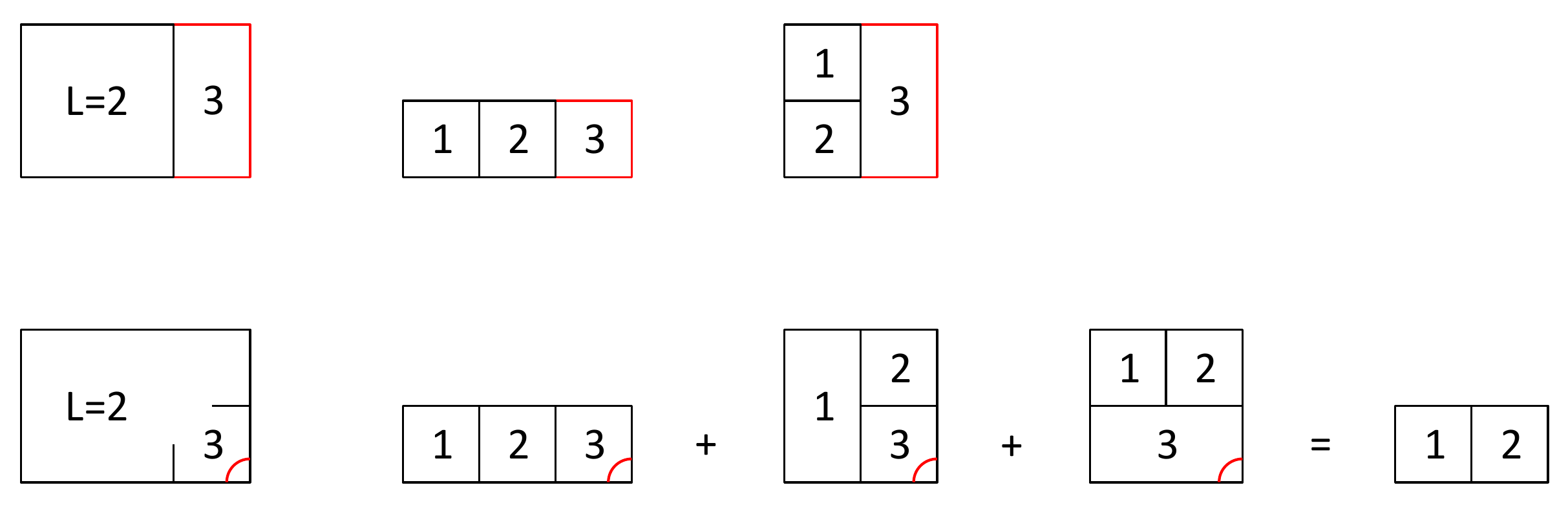}
\caption{Rectangle removal $x_3\!=\!\infty$, $z_3\!=\!0$ and
corner removal $x_3\!=\!y_3\!=\!\infty$, $z_3\!=\!w_3\!=\!0$. Note the rectangle removal is a special (and simplified)
inverse operation of the rung rule \cite{Bern:2006ew}.} \label{fig-1}
\end{center}
\end{figure}

Let's be more concrete and immediately look at the 3-loop case: $x_3\!\to\!\infty$ and $z_3\!=\!0$ lead to
\be
D_{i3}\to\infty\,z_i+(y_3-y_i)(w_i-w_3),
\ee
so $D_{13},D_{23}$ are positive, and we may claim that the third loop ``decouples'' from the rest two loops while
positivity of $D_{12}$ remains to be imposed. Now according to the integrands defined above (in terms of the 3-loop result
$R_3$ given in \cite{Rao:2017fqc}, see figure \ref{fig-0}), namely
\be
R_2=\frac{x_2z_1+x_1z_2+y_2w_1+y_1w_2}{D_{12}},
\ee
\be
\bal
R_3=&\,\frac{x_2x_3z_1z_2+y_2y_3w_1w_2}{D_{12}D_{23}}
+\frac{x_3^2z_1z_2\,y_2w_1+x_2x_3z_1^2\,y_3w_2+x_2z_1\,y_3^2w_1w_2+x_3z_2\,y_2y_3w_1^2}{D_{12}D_{13}D_{23}}\\
&+(5\textrm{ permutations of 1,2,3})
\eal
\ee
for $L\!=\!2,3$, the residue of $R_3$ at $x_3\!=\!\infty,z_3\!=\!0$ is exactly $R_2$! This simple relation reflects
the consistency of 4-particle amplituhedron as expected. We can further make it a bit more nontrivial by similarly setting
$y_3\!=\!\infty,w_3\!=\!0$, then
\be
D_{i3}=\frac{1}{\ep}\,(z_i+w_i),
\ee
where the infinitesimal $\ep$ is used to characterize the divergence of both $x_3$ and $y_3$. Now the same relation
$R_3\!\to\!R_2$ also holds but in a more interesting way as we will explain. Both situations above in fact encode the
\textit{new} Mondrian diagrammatics: in the first case a rectangle-like loop is removed, while in the second a
corner-like loop is removed, as visualized in figure \ref{fig-1}. For the \textit{rectangle removal}, as shown in the 1st
line of figure \ref{fig-1}, two examples of non-vanishing contributions are given, from which the reduction from $R_3$ to
$R_2$ can be transparently seen. The non-vanishing criterion of a diagram is, its third loop must have contacts with the
external faces $y,z,w$ (here we keep using the convention in \cite{Rao:2017fqc,An:2017tbf,Rao:2018uta}, namely faces
$x,z,y,w$ locate at the left, right, top and bottom of the diagram respectively). Similarly for the \textit{corner removal},
a diagram must let its third loop have contacts with the external faces $z,w$ in order to be non-vanishing.
Of course, the rectangle or corner removal can have different choices of orientation by dihedral symmetry
but without loss of generality, we stick to $y,z,w$ and $z,w$ for consistency as above.

However, unlike the rectangle removal for which each 3-loop diagram simply reduces to a 2-loop one, the corner removal
leads to more interesting relations among different 3-loop diagrams, as they together reduce to a 2-loop counterpart.
As shown in the 2nd line of figure \ref{fig-1}, after removing the third loop, these three diagrams reduce to the same
2-loop diagram but with various prefactors, of which the sum is unity:
\be
\frac{z_2}{z_2+w_2}+\frac{z_1}{z_1+w_1}\frac{w_2}{z_2+w_2}+\frac{w_1}{z_1+w_1}\frac{w_2}{z_2+w_2}=1,
\ee
if we define
\be
X_i=\frac{z_i}{z_i+w_i},~~Y_i=\frac{w_i}{z_i+w_i},~~D_i=X_i+Y_i=1, \labell{eq-1}
\ee
this is exactly the \textit{Mondrian completeness relation} \cite{An:2017tbf}, which is trivial to prove:
\be
D_1X_2+X_1Y_2+Y_1Y_2=D_1D_2.
\ee
Diagrammatically Mondrian factors $X_i,Y_i,D_i$ mean loop 3 has a horizontal contact, vertical contact or no contact
with loop $i\!=\!1,2$ respectively. From these easy examples of two types of loop removal, we see how the
Mondrian diagrammatics helps understand the interconnections among different diagrams of various DCI topologies
(including their coefficients) in an extremely neat way.

\newpage
\section{Nontrivialities at 4-loop}

\begin{figure}
\begin{center}
\includegraphics[width=0.75\textwidth]{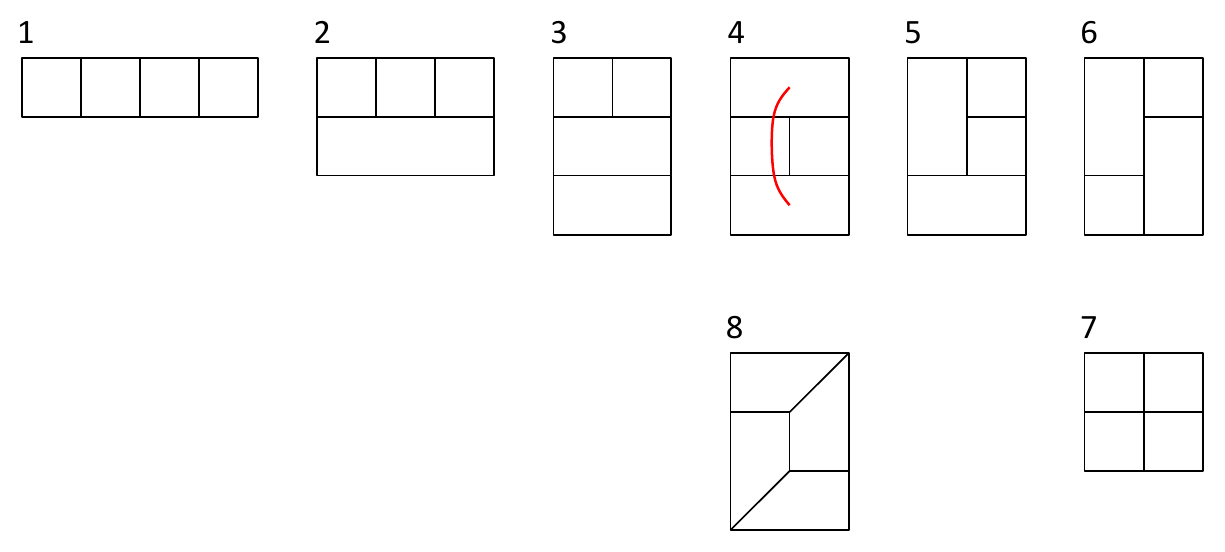}
\caption{4-loop DCI topologies (see \cite{Bern:2006ew}). The red curve denotes a rung rule factor $D_{ij}$,
while rung rule factors $x,y,z,w$ are diagrammatically suppressed for convenience.} \label{fig-2}
\end{center}
\end{figure}

Next, we are curious to see how this wishfully simple mechanism deals with the more sophisticated 4-loop case, since it
involves DCI topologies with coefficient $-1$ and non-Mondrian pole structure (recall that for a
\textit{Mondrian diagram}, all internal lines can be oriented either horizontally or vertically), both of which are
absent at lower loop orders. First of all, let's recall the 4-loop DCI topologies as given in figure \ref{fig-2}.

\begin{figure}
\begin{center}
\includegraphics[width=0.6\textwidth]{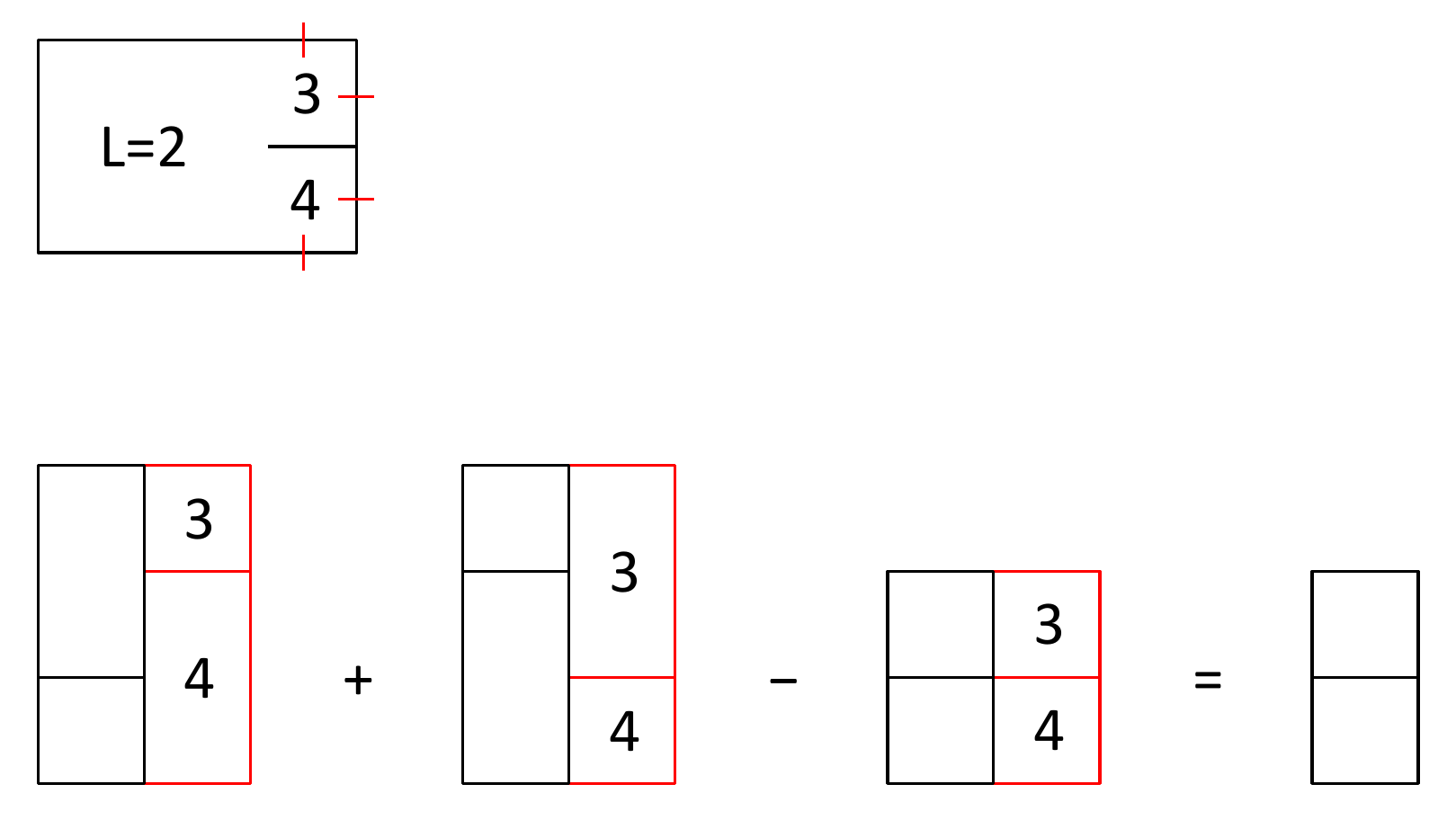}
\caption{Block removal and the cancelation between $T_6$ and $T_7$ in $R_4\!\to\!R_2$.} \label{fig-3}
\end{center}
\end{figure}

Here topologies $T_1,\ldots,T_7$ are Mondrian while $T_8$ is not. As we have known, $T_7$ and $T_8$ are associated with
coefficient $-1$, and $T_1,\ldots,T_6$ are associated with $+1$, moreover, $T_4$ has a $D_{ij}$ factor in the numerator of
its integrand. To see why these coefficients are so, we can pretend that they are still unknown yet and denote them by
$s_1,\ldots,s_8$. Immediately, we can perform the rectangle removal of loop 4. More rigorously speaking, we impose the limit
\be
x_4=\frac{1}{\ep},~~z_4=0~~\textrm{where}~~\ep\to0
\ee
in the 4-loop integrand $R_4$, which takes into account DCI loop integrals of all possible orientations given by dihedral
symmetry (the number of which can be 8, 4, 2, or 1 for each topology) and all $4!$ permutations of loop numbers
\cite{Bern:2006ew}. Then in the expansion
\be
R_4(\ep)=R_4(0)+O(\ep),
\ee
the leading term $R_4(0)$ depends on $s_1,s_2,s_3,s_4,s_5$ only, and $R_4(0)\!=\!R_3$ when
$s_1\!=\!s_2\!=\!s_3\!=\!s_4\!=\!s_5\!=\!1$ as expected. From the Mondrian diagrammatic perspective, this is trivial to
understand as a 4-loop example of the rectangle removal. In fact we can further generalize the rectangle to remove more
loops at a time, which will justify the existence of $T_6,T_7$. As visualized in figure \ref{fig-3}, we now remove a
\textit{block} containing loop $3,4$ by imposing
\be
x_3=x_4=\frac{1}{\ep},~~z_3=z_4=y_3=w_4=0,
\ee
so that for $i\!=\!1,2$
\be
D_{i3}=D_{i4}=\frac{1}{\ep}\,z_i,~~D_{34}=y_4w_3,
\ee
note that $\ep$ helps regularize the expression $(\infty\!-\!\infty)$ in $D_{34}$ and renders this factor vanish.
Then loop $3,4$ decouple from the rest two loops, and in the expansion
\be
R_{4,2}(\ep)=R_{4,2}(0)+O(\ep)
\ee
where the additional subscript 2 of $R_4$ denotes removing two loops at a time, we find
\be
R_{4,2}(0)=\frac{s_3\,(x_2z_1+x_1z_2)+(2\,s_6+s_7)(y_2w_1+y_1w_2)}{D_{12}},
\ee
which equals $R_2$ when $s_3\!=\!s_6\!=\!1$ and $s_7\!=\!-1$. This is also easy to understand diagrammatically, and the
interesting combination $(2s_6\!+\!s_7)$ explains why we need a minus sign for the cross topology $T_7$: the block
removal of loop $3,4$ of two different orientations of $T_6$ gives the same 2-loop diagram, therefore one must be eliminated
in order to maintain $R_{4,2}(0)\!=\!R_2$ while all orientations of $T_6$ and $T_7$ are used exactly once, as shown in
figure \ref{fig-3}. This cancelation mechanism is somehow analogous to the cancelation between the cross and the
brick-wall patterns in Mondrian diagrammatics \cite{An:2017tbf}, and we will see more examples at higher loop orders
reflecting the same essence.

\begin{figure}
\begin{center}
\includegraphics[width=0.6\textwidth]{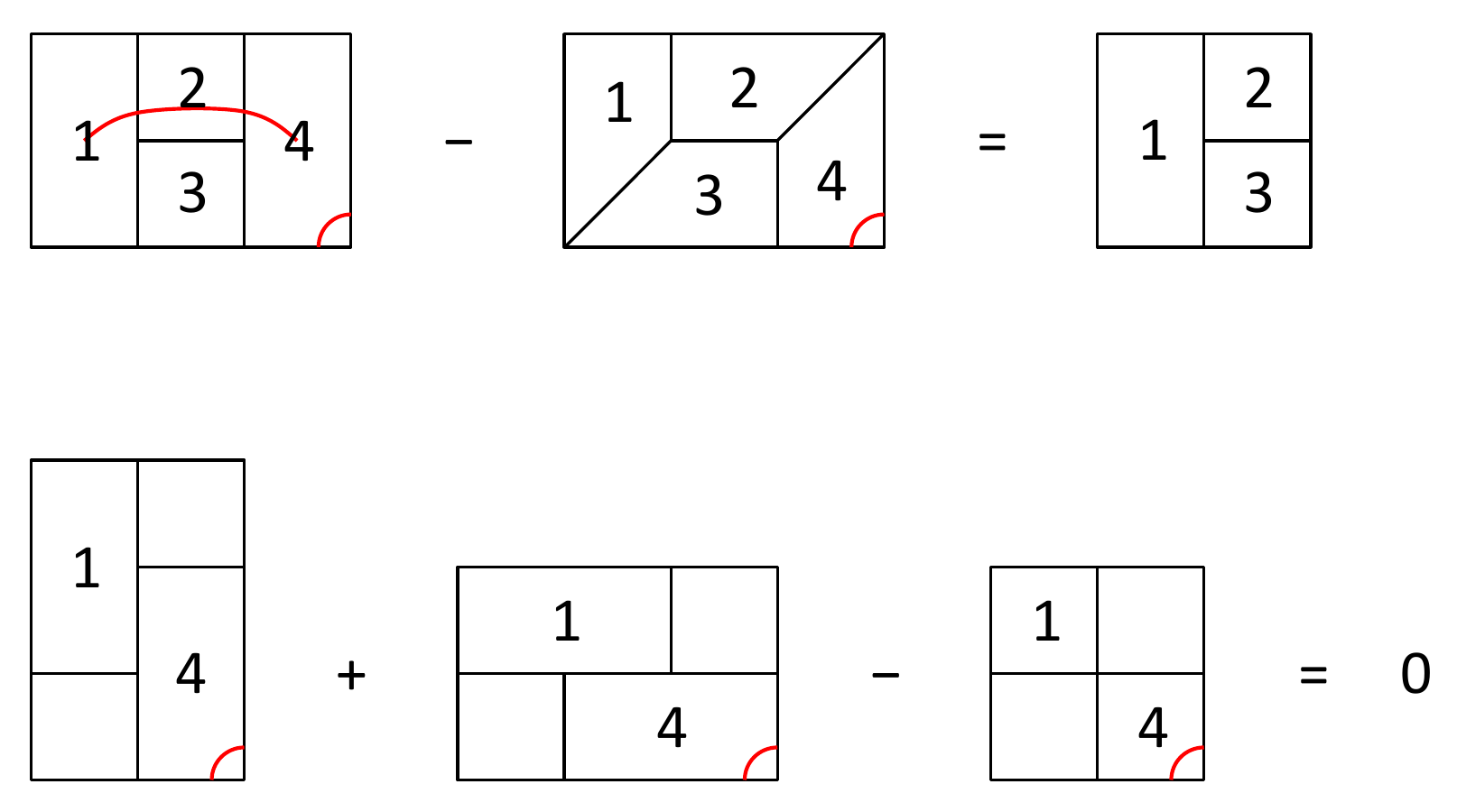}
\caption{New features of the 4-loop corner removal.} \label{fig-4}
\end{center}
\end{figure}

Now only $s_8$ awaits to be explained and we must use the corner removal to detect this non-Mondrian topology $T_8$,
since it has no rectangle or block to be properly removed. Similarly, for removing loop 4 we impose the limit
\be
x_4=y_4=\frac{1}{\ep},~~z_4=w_4=0,
\ee
then in the expansion $R_4(\ep)\!=\!R_4(0)\!+\!O(\ep)$, we find
\be
R_4(0)-R_3\propto1+s_8,
\ee
to maintain the consistency we must take $s_8\!=\!-1$. This is easy to understand if we look at $T_4,T_8$
together among others, as the $D_{ij}$ factor in the numerator of $T_4$'s integrand requires a counter term for producing
the correct Mondrian factor. More concretely, in the 1st line of figure \ref{fig-4}, the relevant two diagrams give
\be
\frac{x_2x_3x_4z_1z_2z_3\,y_3w_2}{D_{12}D_{13}D_{23}D_{24}D_{34}}\,(D_{14}-y_4w_1)
=\frac{z_2}{z_2+w_2}\frac{z_3}{z_3+w_3}\times\frac{x_2x_3z_1^2\,y_3w_2}{D_{12}D_{13}D_{23}}+O(\ep)
\ee
after using
\be
D_{14}-y_4w_1=\frac{1}{\ep}\,(z_1+w_1)-\frac{1}{\ep}\,w_1=\frac{1}{\ep}\,z_1,
\ee
where
\be
\frac{z_2}{z_2+w_2}\frac{z_3}{z_3+w_3}=X_2X_3
\ee
is the desired Mondrian factor, which characterizes the contacting relation between loop 4 in the diagram of topology $T_4$
and its 3-loop sub-diagram (or the resulting 3-loop diagram at the RHS).

Moreover, the cancelation between $T_6$ and $T_7$ in the 4-loop corner removal $R_4\!\to\!R_3$ again holds,
as the relevant three diagrams in the 2nd line of figure \ref{fig-4} lead to the combination
\be
x_4z_1+y_4w_1-D_{14}=\frac{1}{\ep}\,z_1+\frac{1}{\ep}\,w_1-\frac{1}{\ep}\,(z_1+w_1)=0,
\ee
which is exactly isomorphic to the cancelation between the cross and the brick-wall patterns in Mondrian diagrammatics
\cite{An:2017tbf}.

Let's summarize the nontrivialities in the 4-loop case via understanding $s_1,\ldots,s_8$. First, it is useful to
generalize the rectangle removal to the block removal, in order to check the consistency of decoupling more than one loop
at a time. At 6-loop order we will also need the block removal of three loops, and so on.
Next, topologies $T_7,T_8$ with coefficient $-1$ serve as counter terms of those with $+1$, and while $T_7$ has a clear
meaning in Mondrian diagrammatics, $T_8$ appears to be the necessary \textit{company} of $T_4$ which has a nontrivial
$D_{ij}$ factor in its integrand. At 5-loop order and higher, even a company topology or a group of company topologies will
have its further company. While the contributing topologies are more diverse, the overall Mondrian consistency is
maintained by these company topologies.

\section{Rectangle and Block Removals at 5-loop}

\begin{figure}
\begin{center}
\includegraphics[width=1.0\textwidth]{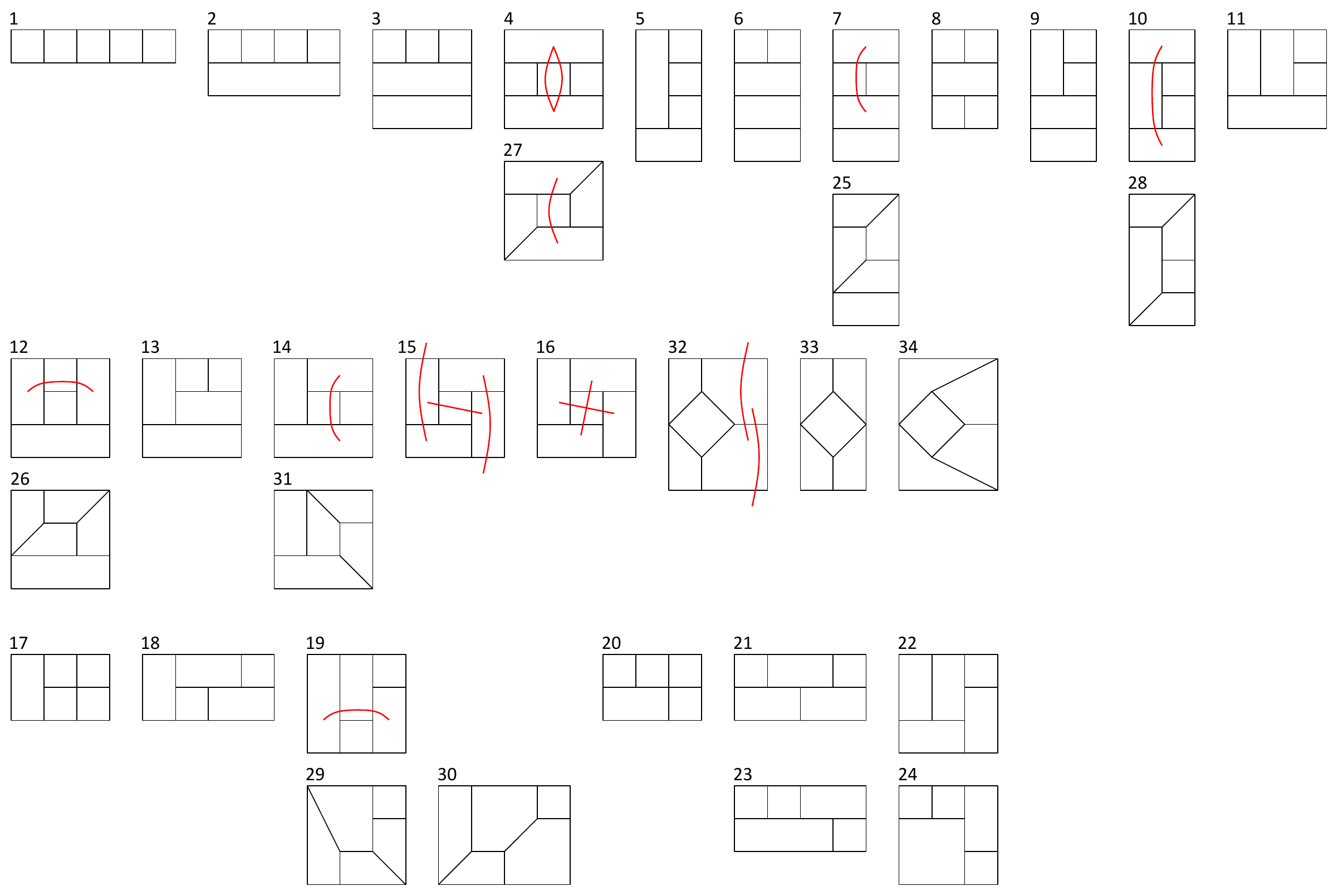}
\caption{5-loop DCI topologies (see \cite{Bern:2007ct}). The red curves denote $D_{ij}$ factors, while $x,y,z,w$ factors
for $T_{15}$ and $T_{32}$ are also indicated for clarity.} \label{fig-5}
\end{center}
\end{figure}

To see more nontrivial examples of various patterns found at $L\!=\!3,4$, and to check whether new features or
exceptions appear, we move on to the 5-loop case. First let's recall the 5-loop DCI topologies as given in
figure \ref{fig-5}, where all 34 topologies are reorganized for later convenience while the same labels used
in \cite{Rao:2018uta} are kept.

Note that according to the classification in \cite{An:2017tbf}, $T_1,\ldots,T_{14}$ are of the ladder type
and $T_{17},\ldots,T_{24}$ are of the cross and brick-wall types. The coefficients or signs of these topologies can be
immediately determined by the rule that each cross pattern contributes $-1$ multiplicatively (otherwise 1),
explicitly we have
\be
s_1=s_2=s_3=s_4=s_5=s_6=s_7=s_8=s_9=s_{10}=s_{11}=s_{12}=s_{13}=s_{14}=1,
\ee
\be
s_{17}=s_{20}=-1,~~s_{18}=s_{19}=s_{21}=s_{22}=s_{23}=s_{24}=1.
\ee
Additionally, each of $T_{25},T_{26}$ has an obvious attached rectangle, therefore their signs automatically follow
that of $T_8$ in the 4-loop case, namely $s_{25}\!=\!s_{26}\!=\!-1$.

Upon these inputs, we find the rectangle removal of loop 5
\be
x_5=\frac{1}{\ep},~~z_5=0
\ee
leads to $R_5(0)\!=\!R_4$ as expected. And the block removal of loop $4,5$
\be
x_4=x_5=\frac{1}{\ep},~~z_4=z_5=y_4=w_5=0
\ee
leads to
\be
R_{5,2}(0)-R_3\propto2\,s_{15}+s_{16}+2\,s_{32}+s_{33},
\ee
which should be zero as required by the consistency. To confirm this guess and to further fully understand the 5-loop case,
let's identify $s_{15},s_{16}$ and $s_{27},\ldots,s_{34}$ one by one, similar to the identification of $s_8$ in the
4-loop case.

\section{Identifications of the Rest Coefficients}

First of all, $s_{28},s_{29},s_{30}$ can be trivially determined by the 4-loop knowledge. Obviously, $T_{28}$ is the
company topology of $T_{10}$, similar to the fact that $T_8$ is the company topology of $T_4$ in the 4-loop case,
as the $T_{10},T_{28}$ pair is the counterpart of the $T_4,T_8$ pair plus one rung. Similarly, $T_{29},T_{30}$ are the
company topologies of $T_{19}$, note that $T_{29}$ has one rung rule factor and $T_{30}$ has two substitution rule factors
which result from the corresponding rung rule factor of $T_{19}$. Therefore we simply
have $s_{28}\!=\!s_{29}\!=\!s_{30}\!=\!-1$.

\begin{figure}
\begin{center}
\includegraphics[width=0.45\textwidth]{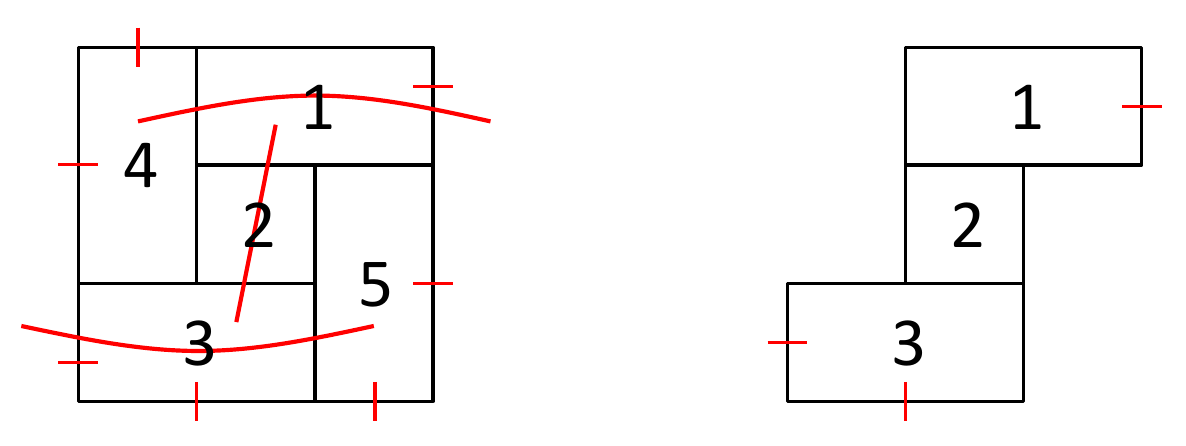}
\caption{Identification of $s_{15},s_{16}$: removing loop 4,5 with additional cuts on loop 1,3.} \label{fig-6}
\end{center}
\end{figure}

Now let's consider $s_{15},s_{16}$ which are a bit tricky. To separate $s_{15},s_{16}$ from other unknown coefficients,
we impose as many cuts as possible around the rim of a particular $T_{15}$ diagram, as shown in figure \ref{fig-6},
and then use a special two-step expansion. To disentangle $s_{15}$ and $s_{16}$, we also relax one external cut,
namely $y_1\!=\!0$ here, and the reason to do so will be clear shortly. Explicitly, we impose the limit
\be
x_5=z_4=\frac{1}{\ep_1},~~y_5=w_4=\frac{1}{\ep_2},~~y_4=x_4=x_3=w_3=w_5=z_5=z_1=0,
\ee
then in the two-step expansion (note the order of expansions matters as we intentionally utilize the $x_5,z_4$ factors
of this diagram to separate it from other sub-leading contributions)
\be
R'_5(\ep_1,\ep_2)=R'_5(0,\ep_2)+O(\ep_1),~~R'_5(0,\ep_2)=R'_5(0,0)+O(\ep_2),
\ee
where the prime denotes additional cuts $x_3\!=\!w_3\!=\!z_1\!=\!0$ besides removing loop 4,5 as indicated
in figure \ref{fig-6}, we find
\be
R'_5(0,0)-R'_3\propto(s_{15}+s_{16})\,y_1w_1-(2\,s_{15}+s_{16}-1)\,y_3w_1-(2\,s_{15}+s_{16}+2\,s_{32}+s_{33})\,x_1z_3.
\ee
To maintain the consistency we must have $s_{15}\!+\!s_{16}\!=\!0$ and $2s_{15}\!+\!s_{16}\!=\!1$, so $s_{15}\!=\!1$ and
$s_{16}\!=\!-1$, which explains why $y_1$ must be non-vanishing, otherwise we cannot identify $s_{15},s_{16}$ with merely
$2s_{15}\!+\!s_{16}\!=\!1$. As we have assumed $2s_{15}\!+\!s_{16}\!+\!2s_{32}\!+\!s_{33}\!=\!0$ in the previous section,
this condition reduces to $2s_{32}\!+\!s_{33}\!=\!-1$ which awaits to be confirmed.

\begin{figure}
\begin{center}
\includegraphics[width=0.17\textwidth]{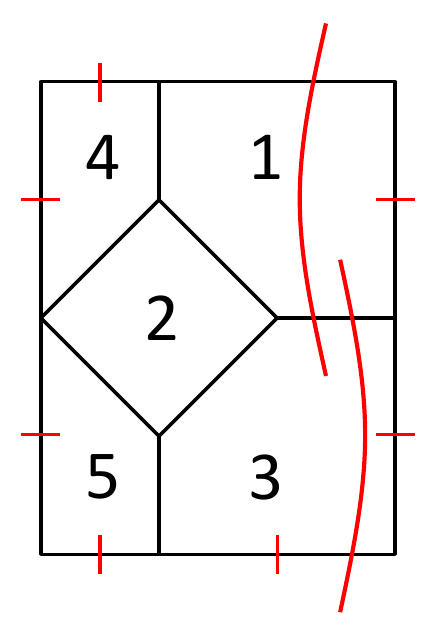}
\caption{Identification of $s_{31},s_{32},s_{33}$: removing loop 4,5 with additional cuts on loop 1,3.} \label{fig-7}
\end{center}
\end{figure}

Next, we can identify $s_{31},s_{32},s_{33}$ upon the inputs of $s_{15},s_{16},s_{30}$ in a similar way.
Picking a particular $T_{32}$ diagram as given in figure \ref{fig-7}, we impose the limit
\be
z_4=w_4=z_5=y_5=\frac{1}{\ep},~~y_4=x_4=x_5=w_5=w_3=z_3=z_1=0,
\ee
and note the external cut of $y_1$ is relaxed, then in the expansion $R'_5(\ep)\!=\!R'_5(0)\!+\!O(\ep)$ we find
\be
R'_5(0)-R'_3\propto(1+s_{31})\,x_2x_3y_1+(s_{32}+s_{33})\,x_2y_1(x_3-w_1)+(1+2\,s_{32}+s_{33})\,x_2(x_1x_3+y_3w_1),
\ee
so the consistency requires $s_{31}\!=\!s_{32}\!=\!-1$ and $s_{33}\!=\!1$. Again, $y_1$ must be non-vanishing, otherwise we
can merely know one condition. Now we have confirmed $2s_{32}\!+\!s_{33}\!=\!-1$ and hence
$2s_{15}\!+\!s_{16}\!+\!2s_{32}\!+\!s_{33}\!=\!0$.

\begin{figure}
\begin{center}
\includegraphics[width=0.17\textwidth]{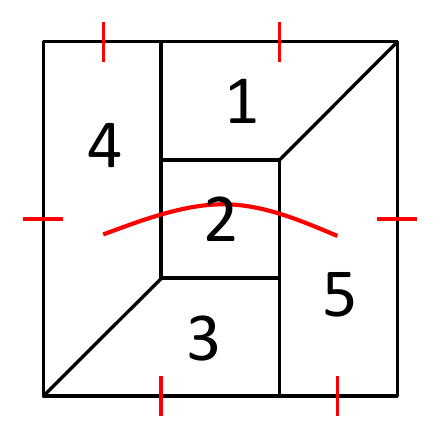}
\caption{Identification of $s_{27}$: removing loop 4,5 with additional cuts on loop 1,3.} \label{fig-8}
\end{center}
\end{figure}

Then for $s_{27}$, we can pick a particular $T_{27}$ diagram as given in figure \ref{fig-8} and impose the limit
\be
z_4=w_4=x_5=y_5=\frac{1}{\ep},~~y_1=y_4=x_4=w_3=w_5=z_5=0,
\ee
now upon the inputs of $s_{15},s_{16}$ and $s_{28},\ldots,s_{33}$, we find
\be
R'_5(0)-R'_3\propto1+s_{27},
\ee
therefore $s_{27}\!=\!-1$.

\begin{figure}
\begin{center}
\includegraphics[width=0.17\textwidth]{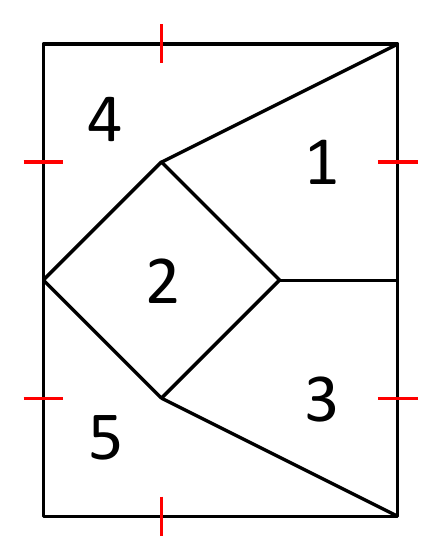}
\caption{Identification of $s_{34}$: removing loop 4,5 with additional cuts on loop 1,3.} \label{fig-9}
\end{center}
\end{figure}

Finally for $s_{34}$, we can pick a particular $T_{34}$ diagram as given in figure \ref{fig-9} and impose the limit
\be
z_4=w_4=z_5=y_5=\frac{1}{\ep},~~y_4=x_4=x_5=w_5=z_3=z_1=0,
\ee
upon the inputs of $s_{15},s_{16}$ and $s_{30},\ldots,s_{33}$ we find
\be
R'_5(0)-R'_3\propto s_{34}-1,
\ee
therefore $s_{34}\!=\!1$. In summary, in this section we have proved that
\be
s_{15}=1,~~s_{16}=-1,~~s_{27}=s_{28}=s_{29}=s_{30}=s_{31}=s_{32}=-1,~~s_{33}=s_{34}=1,
\ee
together with the previous section, all 34 coefficients of 5-loop DCI topologies are now identified.

\newpage
\section{Nontrivial Mondrian Diagrammatic Relations at 5-loop}

Knowing all these coefficients, we then proceed further to understand them, for example, why a coefficient is $-1$
instead of 1, which should not be just an incidental result of imposing cuts at either zero or infinity.
Once we find the Mondrian interconnections among various DCI topologies, their coefficients will become a natural
consequence of these simple relations extracted from the previous derivations.

\begin{figure}
\begin{center}
\includegraphics[width=1.0\textwidth]{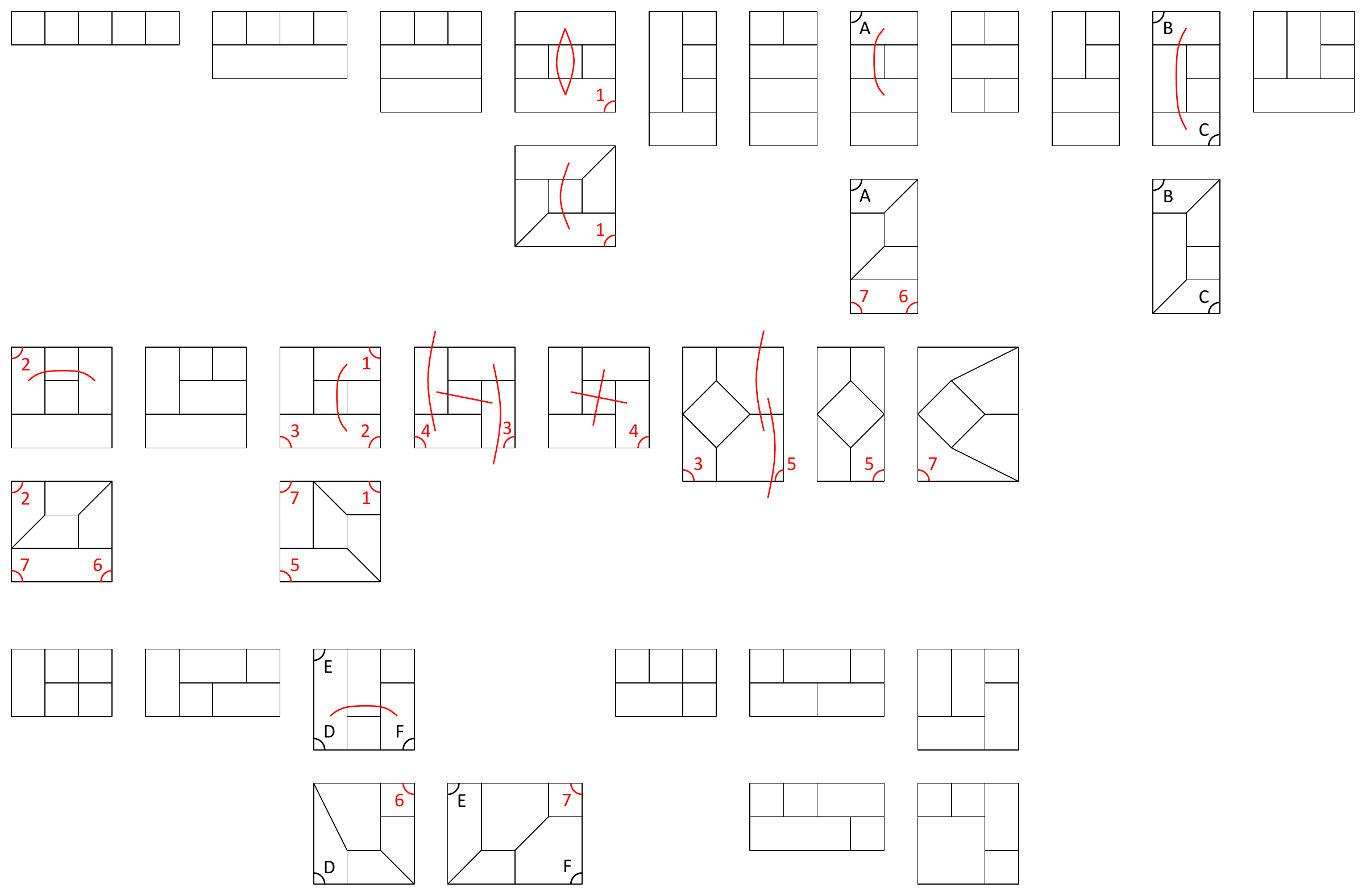}
\caption{Nontrivial corners of 5-loop DCI topologies. Groups $A,\ldots,F$ in black denote relations that are direct
extensions of the 4-loop case. Groups $1,\ldots,7$ in red denote new relations at 5-loop.} \label{fig-10}
\end{center}
\end{figure}

\begin{figure}
\begin{center}
\includegraphics[width=1.0\textwidth]{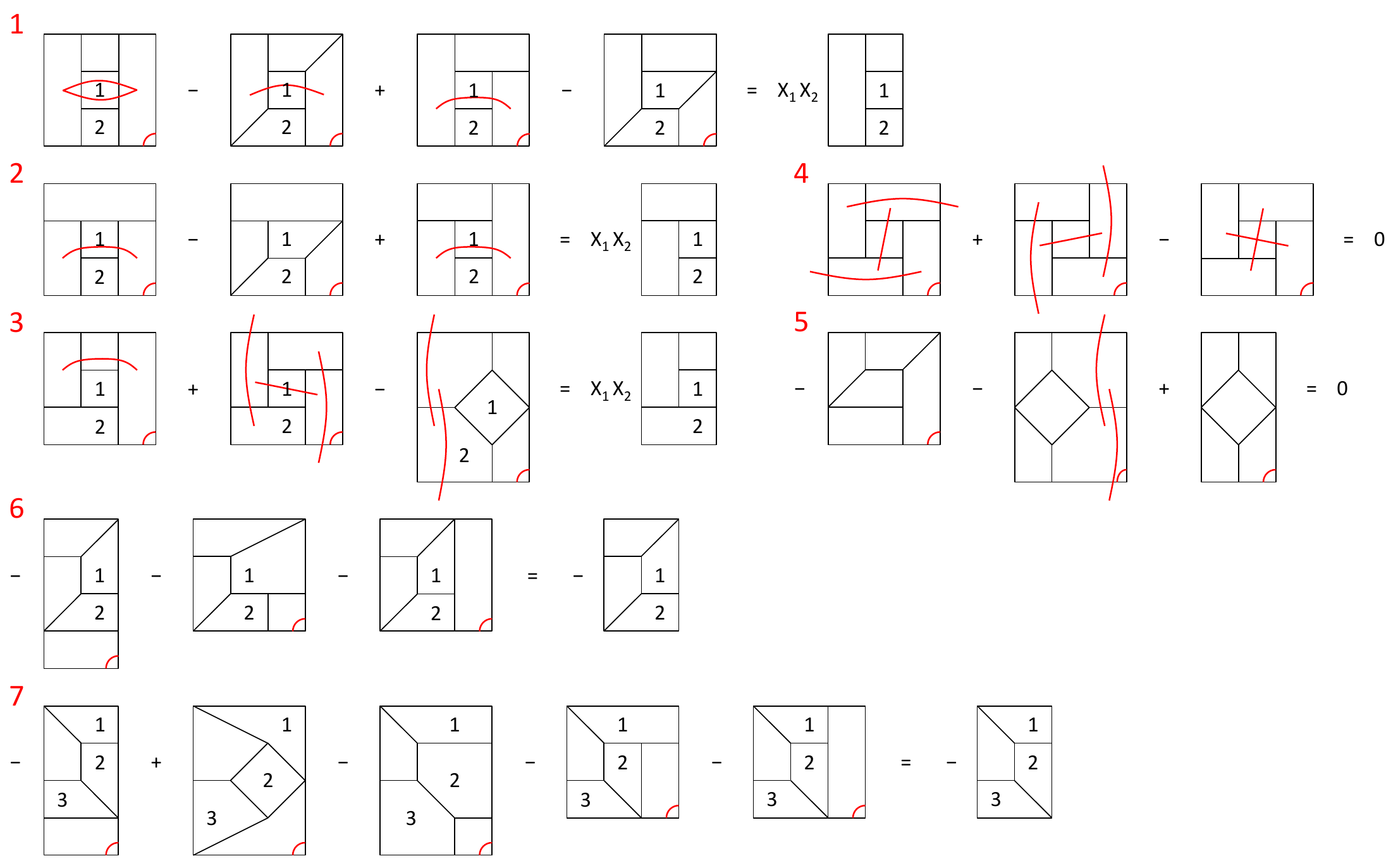}
\caption{New Mondrian diagrammatic relations at 5-loop: groups $1,2,3$ are the relations with obvious Mondrian
pole structures, groups $4,5$ are the vanishing or canceling relations and groups $6,7$ are the relations of
non-Mondrian company topologies.} \label{fig-11}
\end{center}
\end{figure}

More concretely, we would like to explain the universal decoupling relation using one corner removal, which, unlike
the rectangle or block removal, can cover all topologies. The limit to be imposed is simple:
\be
x_5=y_5=\frac{1}{\ep},~~z_5=w_5=0,
\ee
namely removing loop 5, then in the expansion $R_5(\ep)\!=\!R_5(0)\!+\!O(\ep)$ we find $R_5(0)\!=\!R_4$. However,
this is not the end of the story since $R_5(0)\!=\!R_4$ is a redundant relation and it can be further dissected into
many much more transparent sub-relations, as diagrammatically shown in figures \ref{fig-10} and \ref{fig-11}.

In figure \ref{fig-10}, all nontrivial corner removals of 5-loop DCI topologies are indicated. We will focus on groups of
corners denoted by $1,\ldots,7$, while groups denoted by $A,\ldots,F$ are the direct extensions of the corner removals
of $T_4,T_8$ in the 4-loop case (see figure \ref{fig-4}). The rest unmentioned corners are as trivial as the 3-loop
corners or 4-loop corners except those of $T_4,T_8$, as they locate in visually Mondrian topologies and do not
involve the $D_{ij}$ factor.

In figure \ref{fig-11}, groups $1,\ldots,7$ are further separated into three types so that we can more clearly understand
their nontrivialities, let's select one example from each type to elaborate these delicate relations. In the 1st
diagrammatic equality, the 1st and 3rd diagrams have Mondrian pole structures, and in these two diagrams the removed loop has
horizontal contacts with loop 1,2 while its contact with the unlabeled loop on top of loop 1 is horizontal in the
1st diagram and vertical in the 3rd. Naively this should give
\be
X_1X_2X_3+X_1X_2Y_3=X_1X_2D_3=X_1X_2
\ee
according to the definitions in \eqref{eq-1}, which uses a Mondrian completeness relation and here 3 denotes the unlabeled
loop. However, the $D_{ij}$ factors in the 1st and 3rd diagrams complicate this relation and that's why we also need the
2nd and 4th diagrams with minus signs to offset that, then we can exactly get the neat result at the RHS with Mondrian
factor $X_1X_2$. The 2nd and 3rd diagrammatic equalities share the same feature of needing non-Mondrian company topologies,
to offset the extra complexity brought by the $D_{ij}$ factors. However, such a company topology does not have one-to-one
correspondence to a particular Mondrian topology, unlike the $T_4,T_8$ pair at 4-loop.

For the 4th diagrammatic equality, under the corner removal, schematically it is proportional to $X\!+\!Y\!-\!D$,
so it simply vanishes. The 5th equality follows exactly the same cancelation mechanism, though it is not
so obvious as the 4th.

\begin{figure}
\begin{center}
\includegraphics[width=0.35\textwidth]{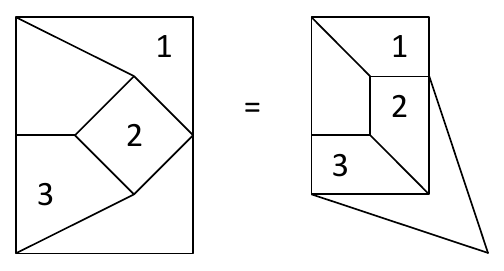}
\caption{Deformation to manifest Mondrian factor $D_1X_2Y_3$.} \label{fig-12}
\end{center}
\end{figure}

For the 6th diagrammatic equality, under the corner removal three non-Mondrian diagrams sum to a 4-loop
non-Mondrian diagram, due to
\be
D_1Y_2+Y_1X_2+X_1X_2=D_1D_2=1.
\ee
Though the resulting 4-loop diagram is not Mondrian, its contact with the 5th loop is still Mondrian, so that we can
use the Mondrian completeness relation. The 7th equality is similar but more nontrivial as
\be
D_1D_2Y_3-D_1X_2Y_3+D_1Y_2X_3+Y_1X_2D_3+X_1X_2D_3=D_1D_2D_3=1,
\ee
note the 2nd diagram has a plus sign so it contributes a minus in the Mondrian completeness relation, as the resulting
4-loop diagram also has a minus sign. The Mondrian factor $D_1X_2Y_3$ from this diagram is not obvious in the
sense of horizontal and vertical contacts, but we can deform its external profile to manifest this,
as shown in figure \ref{fig-12}. Now the external profile of this 5-loop diagram is not a rectangle, but we can see
a familiar 4-loop non-Mondrian diagram hidden in it. With Mondrian factor $D_1X_2Y_3$ clarified, which is the desired
result for offsetting two $D_1X_2Y_3$ factors from the rest four diagrams, we see the Mondrian diagrammatics works
more effectively than naive visual intuition. A final remark is, the 2nd diagram also serves as a company topology of
the rest four, similar to the complexity of the first three identities.

\newpage
\section{Coefficients $+2$ and $0$ at 6-loop}

Finally we can take a glance at the 6-loop case, by investigating the coefficients of two special 6-loop DCI topologies.
First of all, the 6-loop amplituhedron or integrand involves 229 non-vanishing contributions of DCI topologies, among which
125 have Mondrian pole structures as listed in Appendix B of \cite{An:2017tbf}, while the rest 104 non-Mondrian ones are the
company topologies in the sense of corner removal. Interestingly, we also need six vanishing DCI topologies, namely those
with coefficient 0, for a complete understanding of the 6-loop corner removal. All these topologies with coefficients
can be found in the original result \cite{Bourjaily:2011hi}.

\begin{figure}
\begin{center}
\includegraphics[width=0.77\textwidth]{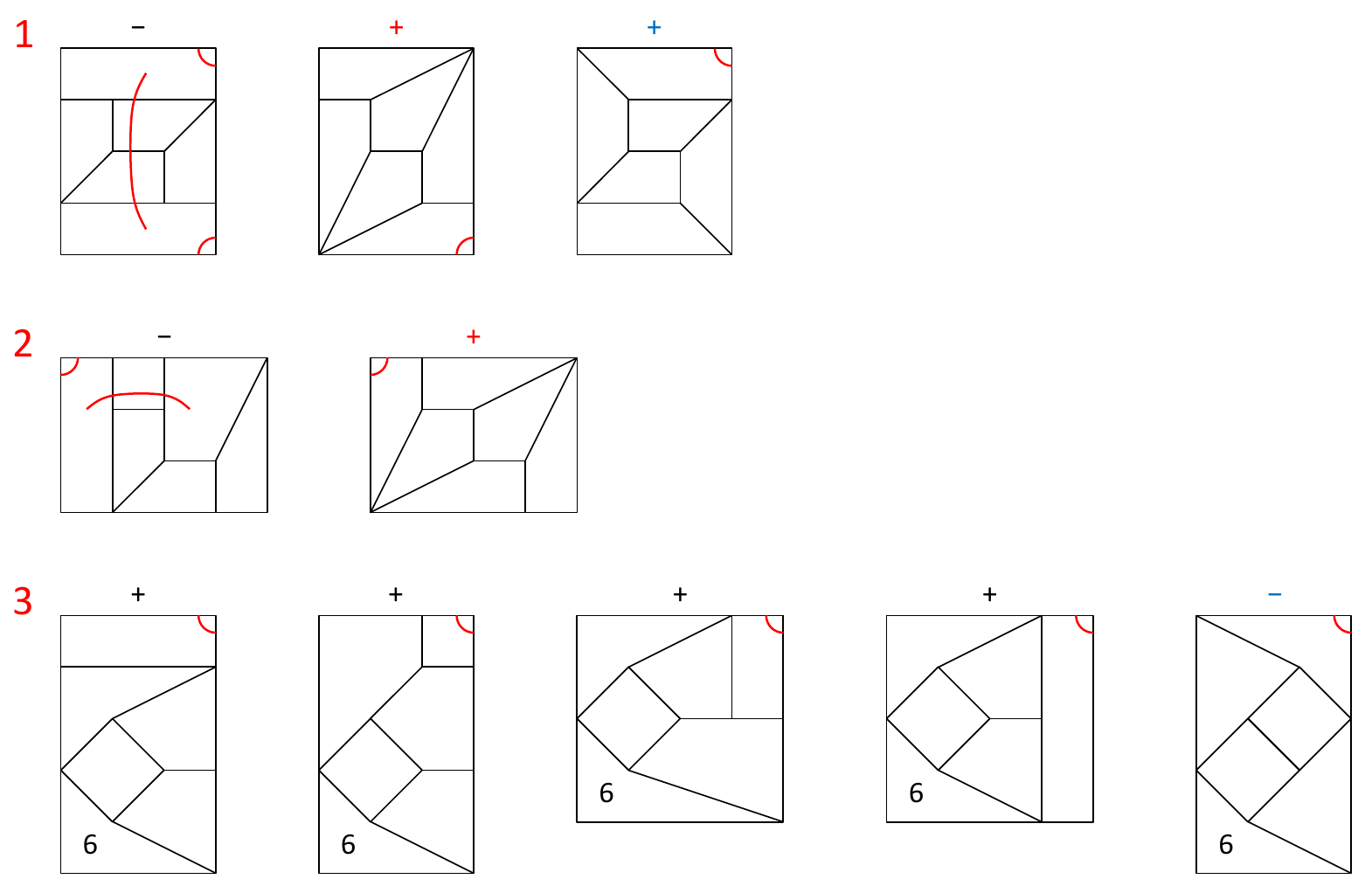}
\caption{The two plus signs in red of the first topology add up to $+2$, while the plus and minus signs in blue
of the second topology add up to 0, in three relevant Mondrian diagrammatic relations.} \label{fig-13}
\end{center}
\end{figure}

Similar to the 5-loop case we have extensively described, the new Mondrian diagrammatic relations at 6-loop also can be
separated into three types: those with obvious Mondrian pole structures, the vanishing or canceling ones and those of
non-Mondrian company topologies. Now we consider two particular 6-loop DCI topologies with three relevant Mondrian
diagrammatic relations, as shown in figure \ref{fig-13}.

The first topology is the 2nd diagram in the 1st relation, or the 2nd diagram in the 2nd relation,
as each of them serves as a company topology of the 1st diagram in the 1st or 2nd relation, following exactly the same
mechanism of $T_4,T_8$ pair at 4-loop (see figure \ref{fig-4}). Since this topology appears twice and in both situations
it has a plus sign, its overall coefficient is simply $+2$ as we add up these two pluses!

Similarly, the second topology is the 3rd diagram in the 1st relation, as a company topology for the other corner of the 1st
diagram, or the 5th diagram in the 3rd relation, and note that they belong to the same topology though drawn differently.
In these two situations it has a plus and a minus respectively, so they cancel and its overall coefficient is 0!
One may find the 3rd relation unfamiliar, but it is simply the 7th relation in figure \ref{fig-11} if we remove the 6th
loop as indicated in figure \ref{fig-13}. Imagine the deformation in figure \ref{fig-12} to better visualize this analogy,
one will find this 6-loop relation completely trivial based on its 5-loop counterpart with an overall sign reverse for all
topologies. Then the two special coefficients $+2$ and 0 are neatly explained, and the 0's of other 6-loop DCI topologies
have similar origins (while there is only one 6-loop DCI topology with coefficient $+2$).

\section{Conclusion and Outlook}

Through substantial examples of DCI integrals at $L\!=\!3,4,5,6$, we discover new interconnections among 4-particle
amplituhedra of different loop orders, the latter are defined by $L(L\!-\!1)/2$ conditions
\be
D_{ij}=(x_j-x_i)(z_i-z_j)+(y_j-y_i)(w_i-w_j)>0,
\ee
which are symmetric up to all loops. For a higher loop amplituhedron, we can use rectangle, block and corner removals
to trivialize $D_{ij}\!>\!0$ conditions involving one or more sets of loop variables $(x_i,y_i,z_i,w_i)$,
so it can reduce to a lower loop one. In this way, we can establish a global view of the all-loop 4-particle
amplituhedron, in particular, we find the coefficients of DCI integrals of different loop orders which seem to be
independent, in fact follow an ``inheriting'' pattern and it can explain many $\pm1$ coefficients. Also at the same loop
order, coefficients of different DCI topologies are constrained by the simple combinatorial rule: the Mondrian
completeness relation. Its underlying mathematics is to repeatedly use $D\!=\!X\!+\!Y$, as $D,X,Y$ characterize the
contacting relations between the removed loop and other loops. To ensure this completeness relation, we also need
non-Mondrian (company) DCI topologies as counter terms.

The integration of the all-loop consistency of 4-particle amplituhedron and Mondrian diagrammatics can explain all
coefficients of DCI topologies at $L\!=\!3,4,5,6$, providing a more transparent supplementary understanding of the results
generated by the soft-collinear bootstrap \cite{Bourjaily:2011hi}. Its simplicity results from the definition of
amplituhedron with its properties under certain limits, and it is an desirable future direction to develop an equally
simple direct construction of the basis of DCI integrals based on this insight.

Historically, since the well known rung rule at 2-loop order \cite{Bern:1997nh}, there have been various rules relating
$L$-loop and $(L\!+\!1)$-loop amplitudes. Besides the algebraic approach \cite{Bourjaily:2011hi} above which imposes the
correct soft-behavior of the logarithm of the amplitude, they also include the correlator-inspired relations used in
\cite{Eden:2012tu} which generalize the rung rule and introduce other rules, and the square plus triangle rules
in \cite{Bourjaily:2016evz}. It is interesting to note that there are some graphical similarities between the square plus
triangle rules and the rectangle (block) plus corner rules. However, mathematically they appear to be quite different
at the current stage of understanding, as the physical meanings of, for example, positive infinity and Mondrian
completeness relation await to be explored. We would like to again emphasize that, at least up to 6-loop, the corner removal
alone can account for all coefficients.

At 7-loop order there is no novelty other than $+2$ and 0 coefficients \cite{Bourjaily:2011hi}, while starting from
the 8-loop case fractional coefficients begin to appear \cite{Bourjaily:2015bpz,Bourjaily:2016evz}. Therefore we expect
a nontrivial generalization of the Mondrian diagrammatic relations at $L\!\geq\!8$, but they should be not too exotic
since these coefficients are still rational. Finally, we would like to explore how the Mondrian consistency connecting
amplituhedra of different loop orders can be extended to the generic case of more than four particles
\cite{Arkani-Hamed:2017vfh,Arkani-Hamed:2018rsk,Kojima:2018qzz}, and what it can tell us about the generic case from the
4-particle knowledge \cite{Heslop:2018zut}.



\begin{thebibliography}{99}


\bibitem{Arkani-Hamed:2013jha}
  N.~Arkani-Hamed and J.~Trnka,
  ``The Amplituhedron,''
  JHEP {\bf 1410}, 030 (2014)
  [arXiv:1312.2007 [hep-th]].


\bibitem{Arkani-Hamed:2013kca}
  N.~Arkani-Hamed and J.~Trnka,
  ``Into the Amplituhedron,''
  JHEP {\bf 1412}, 182 (2014)
  [arXiv:1312.7878 [hep-th]].


\bibitem{Arkani-Hamed:2017vfh}
  N.~Arkani-Hamed, H.~Thomas and J.~Trnka,
  ``Unwinding the Amplituhedron in Binary,''
  JHEP {\bf 1801}, 016 (2018)
  [arXiv:1704.05069 [hep-th]].


\bibitem{Rao:2017fqc}
  J.~Rao,
  ``4-particle Amplituhedron at 3-loop and its Mondrian Diagrammatic Implication,''
  JHEP {\bf 1806}, 038 (2018)
  [arXiv:1712.09990 [hep-th]].


\bibitem{An:2017tbf}
  Y.~An, Y.~Li, Z.~Li and J.~Rao,
  ``All-loop Mondrian Diagrammatics and 4-particle Amplituhedron,''
  JHEP {\bf 1806}, 023 (2018)
  [arXiv:1712.09994 [hep-th]].


\bibitem{Rao:2018uta}
  J.~Rao,
  ``4-particle Amplituhedronics for 3-5 Loops,''
  Nucl.\ Phys.\ B {\bf 943}, 114625 (2019)
  [arXiv:1806.01765 [hep-th]].


\bibitem{Bern:2006ew}
  Z.~Bern, M.~Czakon, L.~J.~Dixon, D.~A.~Kosower and V.~A.~Smirnov,
  ``The Four-Loop Planar Amplitude and Cusp Anomalous Dimension in Maximally Supersymmetric Yang-Mills Theory,''
  Phys.\ Rev.\ D {\bf 75}, 085010 (2007)
  [hep-th/0610248].


\bibitem{Bern:2007ct}
  Z.~Bern, J.~J.~M.~Carrasco, H.~Johansson and D.~A.~Kosower,
  ``Maximally supersymmetric planar Yang-Mills amplitudes at five loops,''
  Phys.\ Rev.\ D {\bf 76}, 125020 (2007)
  [arXiv:0705.1864 [hep-th]].


\bibitem{Bourjaily:2011hi}
  J.~L.~Bourjaily, A.~DiRe, A.~Shaikh, M.~Spradlin and A.~Volovich,
  ``The Soft-Collinear Bootstrap: N=4 Yang-Mills Amplitudes at Six and Seven Loops,''
  JHEP {\bf 1203}, 032 (2012)
  [arXiv:1112.6432 [hep-th]].


\bibitem{Bern:1997nh}
Z.~Bern, J.~Rozowsky and B.~Yan,
``Two loop four gluon amplitudes in N=4 superYang-Mills,''
Phys. Lett. B \textbf{401}, 273-282 (1997)
[arXiv:hep-ph/9702424 [hep-ph]].


\bibitem{Eden:2012tu}
  B.~Eden, P.~Heslop, G.~P.~Korchemsky and E.~Sokatchev,
  ``Constructing the correlation function of four stress-tensor multiplets and the four-particle amplitude in N=4 SYM,''
  Nucl.\ Phys.\ B {\bf 862}, 450 (2012)
  [arXiv:1201.5329 [hep-th]].


\bibitem{Bourjaily:2015bpz}
  J.~L.~Bourjaily, P.~Heslop and V.~V.~Tran,
  ``Perturbation Theory at Eight Loops: Novel Structures and the Breakdown of Manifest Conformality in N=4 Supersymmetric Yang-Mills Theory,''
  Phys.\ Rev.\ Lett.\  {\bf 116}, no. 19, 191602 (2016)
  [arXiv:1512.07912 [hep-th]].


\bibitem{Bourjaily:2016evz}
  J.~L.~Bourjaily, P.~Heslop and V.~V.~Tran,
  ``Amplitudes and Correlators to Ten Loops Using Simple, Graphical Bootstraps,''
  JHEP {\bf 1611}, 125 (2016)
  [arXiv:1609.00007 [hep-th]].


\bibitem{Arkani-Hamed:2018rsk}
  N.~Arkani-Hamed, C.~Langer, A.~Yelleshpur Srikant and J.~Trnka,
  ``Deep Into the Amplituhedron: Amplitude Singularities at All Loops and Legs,''
  Phys.\ Rev.\ Lett.\  {\bf 122}, no. 5, 051601 (2019)
  [arXiv:1810.08208 [hep-th]].


\bibitem{Kojima:2018qzz}
  R.~Kojima,
  ``Triangulation of 2-loop MHV Amplituhedron from Sign Flips,''
  JHEP {\bf 1904}, 085 (2019)
  [arXiv:1812.01822 [hep-th]].


\bibitem{Heslop:2018zut}
  P.~Heslop and V.~V.~Tran,
  ``Multi-particle amplitudes from the four-point correlator in planar $ \mathcal{N} $ = 4 SYM,''
  JHEP {\bf 1807}, 068 (2018)
  [arXiv:1803.11491 [hep-th]].


\end{thebibliography}
\end{document}